\documentclass[sigconf, 10pt]{acmart}
\AtBeginDocument{%
  }

\setcopyright{acmlicensed}
\copyrightyear{2018}
\acmYear{2024}
\acmDOI{XXXXXXX.XXXXXXX}

\usepackage{tcolorbox}
\usepackage{makecell}
\usepackage{xcolor}
\usepackage{soul}
\usepackage{colortbl}
\usepackage{graphicx}
\usepackage{subfigure}
\usepackage{subcaption}
\usepackage{cleveref}
\usepackage{adjustbox}
\usepackage{graphicx}
\usepackage{caption}
\usepackage{subcaption}
\usepackage{booktabs}
\usepackage{multirow}
\usepackage{caption}
\usepackage{amsmath}

\definecolor{mygreen}{RGB}{0,128,0}
\definecolor{myred}{RGB}{255,0,0}
\begin{document}\sloppy

\title{AutoJournaling: A Context-Aware Journaling System Leveraging MLLMs on Smartphone Screenshots}





\author{Tianyi Zhang}
\email{t.zhang59@student.unimelb.edu.au}
\orcid{0000-0002-0778-8844}
\affiliation{%
  \institution{University of Melbourne}
  \city{Melbourne}
  \country{Australia}
}

\author{Shiquan Zhang}
\email{shiquan.zhang@student.unimelb.edu.au}
\orcid{0000-0003-3747-0842}
\affiliation{%
  \institution{University of Melbourne}
  \city{Melbourne}
  \country{Australia}
}

\author{Le Fang} 
\email{le.fang2@unimelb.edu.au}
\orcid{0009-0004-2591-7584}
\affiliation{%
  \institution{University of Melbourne}
  \city{Melbourne}
  \country{Australia}
}

\author{Hong Jia}
\email{hong.jia@unimelb.edu.au}
\orcid{0000-0002-6047-4158}
\affiliation{%
  \institution{University of Melbourne}
  \city{Melbourne}
  \country{Australia}
}

\author{Vassilis Kostakos}
\email{vassilis.kostakos@unimelb.edu.au}
\orcid{0000-0003-2804-6038}
\affiliation{%
  \institution{University of Melbourne}
  \city{Melbourne}
  \country{Australia}
}

\author{Simon D'Alfonso}
\email{dalfonso@unimelb.edu.au}
\orcid{0000-0001-7407-8730}
\affiliation{%
  \institution{University of Melbourne}
  \city{Melbourne}
  \country{Australia}
}

\renewcommand{\shortauthors}{Zhang et al.}

\begin{abstract}
Journaling offers significant benefits, including fostering self-reflection, enhancing writing skills, and aiding in mood monitoring. However, many people abandon the practice because traditional journaling is time-consuming, and detailed life events may be overlooked if not recorded promptly. Given that smartphones are the most widely used devices for entertainment, work, and socialization, they present an ideal platform for innovative approaches to journaling. Despite their ubiquity, the potential of using digital phenotyping—a method of unobtrusively collecting data from digital devices to gain insights into psychological and behavioral patterns—for automated journal generation has been largely underexplored. In this study, we propose AutoJournaling, the first-of-its-kind system that automatically generates journals by collecting and analyzing screenshots from smartphones. This system captures life events and corresponding emotions, offering a novel approach to digital phenotyping.  We evaluated AutoJournaling by collecting screenshots every 3 seconds from three students over five days, demonstrating its feasibility and accuracy. AutoJournaling is the first framework to utilize seamlessly collected screenshots for journal generation, providing new insights into psychological states through digital phenotyping.

\end{abstract}






\maketitle

\section{Introduction}
Journaling is an important way for individuals to document their activities, reflect on themselves, and express their emotions and feelings. People engage in journaling for various reasons, including self-improvement \citep{hensley2020power}, therapy \citep{miller2014interactive}, and writing practice \citep{baresh2022developing}. However, the benefits of journaling can be difficult to quantify in a timely manner, especially given the significant time and effort individuals invest in the process. People often feel that maintaining a journaling habit is challenging when it requires substantial time and effort at the end of the day or over a period to recall events and emotions \citep{hayman2012journaling}. In this context, digital devices like smartphones, with their prevalence and ubiquitous nature, can play a crucial role in assisting with the logging of daily activities.

Smartphones serve as an effective tool for recording daily events into diary form through activity recognition \citep{reyes2016transition, ouyang2022cosmo} due to their constant presence and integral role in daily communication and information retrieval. Their ubiquitous use facilitates not only active user interaction (e.g. screen unlocks, application usage, and communication patterns) but also the passive collection of hardware sensor data, such as GPS location, accelerometer readings, and ambient light levels. This comprehensive data collection enables the analysis, prediction, and monitoring of individuals' mental health and well-being, a process known as digital phenotyping.

Several techniques have been developed to assist journaling on smartphones. For instance, \cite{zhang2016examining} introduced diary in-situ reminders which are triggered when the screen is unlocked. Another tool, DiaryMate \citep{kim2024diarymate}, enhances the journaling experience by using human-AI collaboration (HAI) to generate the next sentence based on the previous sentence. Chatbot assistants have also been extensively explored in prior studies. For example, MindfulDiary \citep{kim2024mindfuldiary}, a question-answer-based chatbot assistant, was developed to aid psychiatric patients in journaling using large language models. Additionally, \citep{angenius2022interactive} explored the conversion of probing into language for conversational interactions. Another study compared the effectiveness of humanoid robots and voice assistants in helping students write journals, finding that humanoid robots were more effective in improving mood \citep{sayis2024technology}. In a digital phenotyping study, a contextual journaling application called MindScape \citep{nepal2024contextual} was introduced to collect behavioral habits via smarpthones and suggest content or provide instructions for users. The activities traced by MindScape include physical movement, sleep pattern, social interaction, screen usage, location and conversation lengths. These were prompted as inputs for large language models \citep{cai2023federated} to personalize journaling interest, meant to remind and suggest significant events in daily life that may affect emotional wellbeing of individuals. While these techniques can assist with prompt generation and event recommendations in the journaling process, the time and effort required to recall and document events can still be significant, potentially leading to incomplete or lost logs if adequate time is not devoted to journaling. In contrast, our work is the first to fully and passively automate the journal writing process without requiring any human effort.


In this study, we introduce AutoJournaling, a novel context-aware journaling system that leverages multi-modality large language models (MLLMs) and screenshots to automatically generate journals, entirely removing the need for human effort. The system is designed around two core tasks: generating factual statements and analyzing perceptions. We structured the framework with both dimensionality reduction (image-to-text-to-journal streamline) and dimensionality expansion (image-to-video-to-journal). To demonstrate its capabilities, we collected continuous screenshots every three seconds from three university students at the University of X over five days and processed the data using AutoJournaling to generate daily journals. We evaluated the system's performance by comparing its outputs with manually produced journals, demonstrating its feasibility. AutoJournaling has the potential to significantly streamline the daily journaling process, improve the monitoring and prediction of psychological well-being, and offer valuable insights for personal reflection and mental health.


\section{Method and Experiments}

\subsection{AutoJournaling Framework}

The AutoJournaling framework collects screenshots from smartphones at adjustable intervals. We designed the AutoJournaling framework to generate journals from continuously collected screenshots using two methods: journal generation from individual screenshots (text-based) and journal generation from concatenated screenshots (video-based) as illustrated in Figure \ref{AutoJouranling-workflows}. The extraction of information from text-based screenshots into fine-grained activity descriptions involves a process of dimensional reduction, translating visual data into detailed text and then summarizing journal events. Conversely, converting images into a video format is a form of dimensional aggregation. We compare the text generation from both video and text formats to evaluate the trade-off between performance and efficiency. For all MLLMs and Large Language Models (LLMs), we consistently used Gemini 1.5 Pro \citep{reid2024gemini} configured with a temperature setting of zero and a $top\_p$ parameter setting of 1.

In the video stream of the AutoJournaling framework, the Frames Per Second (FPS) is set to 30. The prompts used to guide the MLLMs in summarizing events and emotions are depicted in Figure \ref{video_prompt}. In the text stream, images are chunked into sets based on either the size limit or the number of images that the MLLM can process, with the prompts shown in Figure \ref{activity_description_prompt} for detailed event description and Figure \ref{image_prompt} for diary summarization.
\vspace{-4pt}

\captionsetup{font=footnotesize, labelformat=simple, labelsep=colon, labelfont=bf}
\begin{figure}[htbp]
  \centering
  \subfigure{\includegraphics[trim=0pt 0pt 0pt 0pt, clip, width=\linewidth]{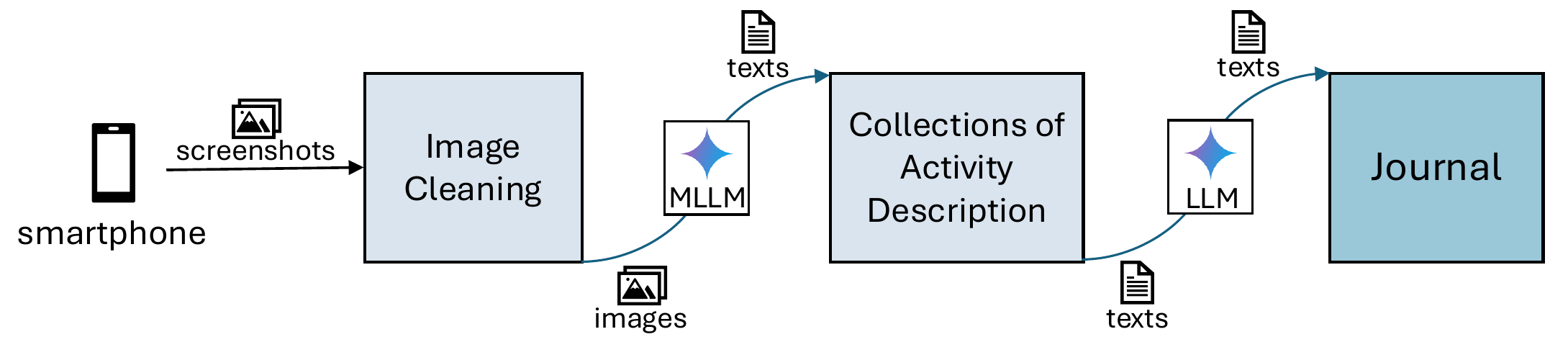}} \label{image_workflow} \\
  \vspace{-18pt}
    \caption*{(a) The process of using LLMs to convert images into text descriptions in small sets, followed by journal generation}
    \vspace{2pt}
  \subfigure{\includegraphics[trim=0pt 0pt 0pt 0pt, clip, width=\linewidth]{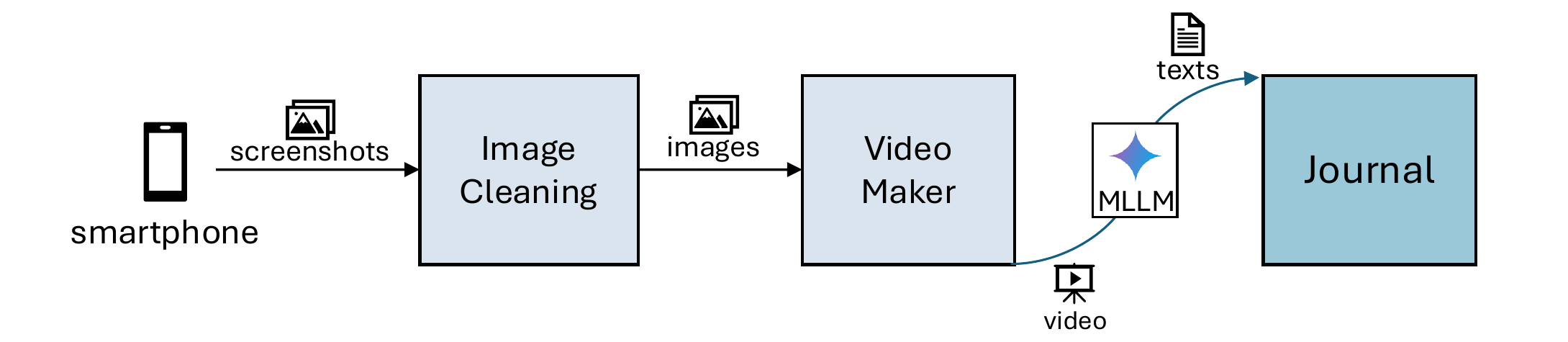}} \label{vidoe_workflow}
  \vspace{-18pt}
    \caption*{(b) The process of converting images into videos, followed by journal generation using LLMs.}
    \vspace{2pt}
  \vspace{-4pt}
  \caption{The AutoJournaling framework workflow}
\end{figure} \label{AutoJouranling-workflows}


We prompt the LLMs to generate journals in an event-based format rather than the traditional continuous narrative format. While journals are typically written as flowing narratives, our approach emphasizes capturing activities through smartphones and inferring the associated emotions. This method is designed to help individuals regulate their mood based on daily behaviors. The event-based format offers a clearer framework for evaluating activities and detecting emotions, providing practical insights for mood monitoring and facilitating the generation of daily reports.


\begin{figure}[htbp]
\centering
\fbox{%
    \parbox{1\columnwidth}{
    \fontsize{8}{8}\selectfont
We took a screenshot of a user's smartphone every \{INTERVAL\} seconds, and the screenshots from \{start\_time\} to \{end\_time\} are attached. Please consider these screenshots as consecutive actions and summarize the user's main activities, highlighting any significant events and the content they were exposed to.
}}
\vspace{-6pt}
\caption{Text analyzer by streams of image sets}
\label{activity_description_prompt}
\end{figure}

\vspace{-8pt}

\begin{figure}[htbp]
\centering
\fbox{%
    \parbox{1\columnwidth}{
    \fontsize{8}{8}\selectfont
    \textsf{\textbf{Instruction}: You are a smartphone behavioral analyst for activity summarization and feeling analysis.} \\

\textsf{\textbf{Question}:} \\
\textsf{Your task is to summarize main events into a diary format, where each entry should be a brief description of the activities, and a speculation about the user's feelings. The events should be significant events, not necessarily with details.} \\

\textsf{\textbf{Context}:}  \\
\textsf{For each entry, three elements should be included:}  \\

\textsf{- events: A brief description of the activities. Focus on the major aspects of events, similar to how one would write in a personal diary. Minor details can be excluded. If any other person is involved in the event, the person and content should be logged. }\\
\textsf{- feelings: A few keywords describing the user's feelings impacted by the event.}\\
\textsf{- reasoning: A brief explanation for the speculated feelings.} \\

\textsf{Additionally, the input data may contain some hallucinated results from the language model; ensure that this content is removed from your output.} \\

\textbf{IMPORTANT:} \\
\textsf{- Avoid repeating the same content (hallucination) in your output.}\\
\textsf{- Each event should have a unique key in the output, arranged chronologically from the first to the last.}\\
\textsf{- Each event should be identical.}\\
\textsf{- The number of the output events should be no more than 30.} \\

\textsf{\textbf{Output Indicator}:}  \\
\textsf{The output should be in the following JSON format:} \\
\{
\textsf{"1": \{"event": <event summary>, "feelings": <speculated feeling>, "reasoning": <reasoning>}\}, \\
\textsf{"2": \{"event": <event summary>, "feelings": <speculated feeling>, "reasoning": <reasoning>}\}, \\
\textsf{...}\}

}}
\captionsetup{font=footnotesize, labelformat=simple, labelsep=colon, labelfont=bf}
\vspace{-6pt}
\caption{Prompt for the image-to-text streamline of the AutoJournaling}
\label{image_prompt}
\end{figure}

\begin{figure}[htbp]
\centering
\fbox{%
    \parbox{1\columnwidth}{
    \fontsize{8}{8}\selectfont
\textsf{\textbf{Instruction:} You are a smartphone behavioral analyst for activity summarization and feeling analysis.} \\

\textsf{\textbf{Question:}} \\
\textsf{The provided video is a capture of screenshots taken every 3 seconds from a smartphone user. Your task is to summarize the main events into a diary format, with each entry being a brief description of the activities and a speculation about the user's feelings. The events should be significant, not necessarily with details.} \\

\textsf{\textbf{Context:}} \\
\textsf{For each entry, three elements should be included:}

\textsf{- events: A brief description of the activities. Focus on the major aspects of events, similar to how one would write in a personal diary. Minor details can be excluded. If any other person is involved in the event, the person and content should be logged.} \\
\textsf{- feelings: A few keywords describing the user's feelings impacted by the event.} \\
\textsf{- reasoning: A brief explanation for the speculated feelings.} \\

\textsf{Additionally, the input data may contain some hallucinated results from the language model; ensure that this content is removed from your output.} \\

\textsf{\textbf{IMPORTANT:}} \\
\textsf{- Avoid repeating the same content (hallucination) in your output.} \\
\textsf{- Each event should have a unique key in the output, arranged chronologically from the first to the last.} \\
\textsf{- Each event should be distinct.} \\
\textsf{- The number of output events should be no more than 30.} \\

\textsf{\textbf{Output Indicator:}} \\
\textsf{The output should be in the following format:} \\
\textsf{\{ "1": \{"event": \textless event summary\textgreater, "feelings": \textless speculated feeling\textgreater, "reasoning": \textless reasoning\textgreater\},} \\
\textsf{"2": \{"event": \textless event summary\textgreater, "feelings": \textless speculated feeling\textgreater, "reasoning": \textless reasoning\textgreater\}, }\\
\textsf{...\} }
}}
\captionsetup{font=footnotesize, labelformat=simple, labelsep=colon, labelfont=bf}
\vspace{-6pt}
\caption{Prompt for the video streamline of the AutoJournaling}
\label{video_prompt}
\end{figure}

\vspace{-8pt}

\subsection{Data Collection and Preprocessing}

For data collection, we utilized the AWARE-Light \citep{van2023aware} smartphone sensing platform, which includes functionality to passively collect screenshots. To demonstrate the feasibility of the AutoJournaling framework, we collected screenshot data from three university students over a five-day period. The screenshots were captured every three seconds. Participants were also instructed to write daily journals, documenting their life events and corresponding feelings.

Before processing, the raw screenshots undergo a preprocessing stage to remove duplicates and invalid images. To optimize time efficiency of analysis while retaining the same level of information, we compare the pixels of the images and calculate a similarity score. The threshold for this score is set to the maximum value of one, meaning that any duplicate images detected as exactly the same will be discarded. In the video making, the generation of the video will not adhere to a strict three-second interval per frame. We expect that LLMs will recognize the time displayed on the smartphone screen, rather than relying on the progress bar of the video, for accurate timestamping.

The human-produced journals were preprocessed to remove events unrelated to smartphone usage for evaluation purposes. Any information that could not be inferred from smartphone data was excluded from the evaluation. For instance, if participants encountered difficulties while working on their laptops, this information might not be captured on their smartphones unless explicitly mentioned. In such cases, this information was discarded during the preprocessing of the human-produced journals. This ensures that the comparison focused solely on events and activities that could reasonably be captured through the smartphone screenshots. 
\vspace{-6pt}
\subsection{Evaluation Framework}
The human-generated journals are preprocessed into a clean JSON format, including both factual statements and psychological states. The journals generated by the AutoJournaling system will follow the same structure. We utilize the Bidirectional Encoder Representations from Transformers (BERT) model to calculate similarity scores for both the factual and psychological sections by comparing the semantic content of event descriptions, which computes semantic similarity between sentences while understanding their context \citep{reimers2019sentence}. For each individual event or each individual psychological statement, the similarity score ranges from -1 to 1, where a score near 1 indicates high similarity, a score near 0 indicates unrelated content, and a score near -1 indicates direct opposites.


The event similarity scores are computed to evaluate how closely the predicted events align with the human-produced ground truth events. Given that the number of events in the ground-truth journal may vary, the similarity score for each AutoJournaling-generated event is defined as the maximum score when compared with all events in the ground-truth journal. For each ground truth event $T_i$, we find the predicted AutoJournaling event $P_j$ with the maximum similarity score. This score is then averaged over all ground truth events to calculate the overall similarity score for the ground truth events, as shown in Equation \eqref{eq:event_score_t}. Similarly, for each predicted event $P_j$, we find the ground truth event $T_i$ with the maximum similarity score, which is then averaged to get the overall similarity score for the predicted events, as shown in Equation \eqref{eq:event_score_p}.


\vspace{-8pt}
\begin{align}
\small
\text{Score}_T^{\text{Event}} &= \frac{1}{n} \sum_{i=1}^{n} \max_{j \in \{1, 2, \dots, m\}} S(T_i, P_j) \label{eq:event_score_t} \\
\small
\text{Score}_P^{\text{Event}} &= \frac{1}{m} \sum_{j=1}^{m} \max_{i \in \{1, 2, \dots, n\}} S(T_i, P_j) \label{eq:event_score_p}
\end{align}
\vspace{-8pt}

The evaluation of feeling similarity scores follows a similar methodology to the calculation of event similarity scores. However, the feeling scores are not assessed independently of the events. First, the events are matched based on the maximum similarity scores. Then, the feeling scores of the corresponding matched events are calculated. Finally, the average of these feeling scores is computed for each set of events, as shown in Equations \ref{eq:feeling_score_t} and \ref{eq:feeling_score_p}.


\begin{align}
\small
\text{Score}_T^{\text{Feeling}} &= \frac{1}{n} \sum_{i=1}^{n} S_F(T_i^{\text{feeling}}, P_{j^*(i)}^{\text{feeling}}) \label{eq:feeling_score_t} \\
\small
\text{Score}_P^{\text{Feeling}} &= \frac{1}{m} \sum_{j=1}^{m} S_F(T_{i^*(j)}^{\text{feeling}}, P_j^{\text{feeling}}) \label{eq:feeling_score_p}
\end{align}

where 

\begin{equation}
\small
i^* = \arg\max_{i \in \{1, 2, \dots, n\}} S(T_i, P_j) \label{eq:i_star}
\end{equation}

\begin{equation}
\small
j^* = \arg\max_{j \in \{1, 2, \dots, m\}} S(T_i, P_j) \label{eq:j_star}
\end{equation}

The evaluation score for ground truth ($Score_T$) measures the extent to which the information in the human-written journals is reflected in the predicted content, similar to the concept of false negatives or negative residuals. The evaluation score for predictions ($Score_P$) indicates the degree to which the predicted content matches the actual content in the ground truth, akin to false positives or positive residuals. The overall score is then computed using the harmonic mean of these two scores for events and feelings respectively, as described by the adapted F-score in Equation \ref{F1}.

\begin{equation}
\small
\text{Score} = \frac{2 \times (score_T \times score_P)}{score_T + score_P} \label{F1}
\end{equation}





\section{Results \& Discussion}
In this section, we discuss the system performance (\autoref{performance}) and result cases (\autoref{case_study}) obtained from the data of the three participants as well as the privacy and ethical issues that may arise (\autoref{privacy}).

\subsection{Performance of the AutoJournaling system} \label{performance}
The results primarily demonstrate the feasibility of using the AutoJournaling system to accurately log both factual events and the associated psychological feelings. Both the text-based (Table \ref{image_performance}) and video-based (Table \ref{video_performance}) streams produce high similarity scores in event descriptions and feeling speculations, demonstrating the system's ability to accurately capture and describe smartphone-related activities and corresponding psycholgoical analysis with three-second screenshots.

Comparing the text-based and video-based stream, no significant differences in system performance were observed. The text-based stream produces detailed descriptions of events throughout the process, while the video-based stream directly outputs summarized events and associated feelings. The processing time for both approaches varies depending on smartphone usage duration. Our findings indicate that the more time users spend on their smartphones, the longer the processing time required for the text-based stream, whereas the video-based stream proves to be more time-efficient under these conditions. At this stage, the AutoJournaling system can serve as a valuable tool for mental health and mood monitoring, offering detailed logs of activities conducted via smartphones for both individuals with mental health concerns and those interested in general well-being.

\begin{table}[ht]
\captionsetup{font=footnotesize, labelformat=simple, labelsep=colon, labelfont=bf}
\caption{The performance of journals generated by AutoJournaling from a text-based stream of screenshot images was evaluated using BERT.}\label{image_performance}
\vspace{-6pt}
\resizebox{0.7\linewidth}{!} {
\begin{tabular}{ccccccc}\toprule
\rowcolor{gray!30}
 & \multicolumn{2}{c}{Alice} & \multicolumn{2}{c}{Bob} & \multicolumn{2}{c}{Chris} \\
Day & event & feeling  & event & feeling & event & feeling \\
1 & 0.90 &  0.90  & 0.91 & 0.96 & 0.91 & 0.93 \\
2 & 0.91 &  0.93  & 0.89 & 0.95 & 0.91 & 0.94 \\
3 & 0.91 &  0.91  & 0.90 & 0.96 & 0.90 & 0.94 \\
4 & 0.89 &  0.93  & 0.87 & 0.95 & 0.91 & 0.94 \\
5 & 0.89 &  0.95  & 0.93 & 0.96 & 0.89 & 0.92 \\ \bottomrule
\end{tabular}}
\end{table}

\vspace{-8pt}

\begin{table}[ht]
\captionsetup{font=footnotesize, labelformat=simple, labelsep=colon, labelfont=bf}
\caption{The performance of journals generated by AutoJournaling from a video-based stream of screenshot images was evaluated using BERT.}\label{video_performance}
\vspace{-6pt}
\resizebox{0.7\linewidth}{!} {
\begin{tabular}{ccccccc}\toprule
\rowcolor{gray!30}
& \multicolumn{2}{c}{Alice} & \multicolumn{2}{c}{Bob} & \multicolumn{2}{c}{Chris} \\
Day & event & feeling  & event & feeling & event & feeling \\
1 & 0.91 &  0.96  & 0.92 & 0.95 & 0.92 & 0.95 \\
2 & 0.91 &  0.93  & 0.91 & 0.95 & 0.91 & 0.95 \\
3 & 0.92 &  0.94  & 0.91 & 0.93 & 0.91 & 0.93 \\
4 & 0.90 &  0.94  & 0.87 & 0.91 & 0.87 & 0.91 \\
5 & 0.91 &  0.94  & 0.93 & 0.97 & 0.93 & 0.97 \\ \bottomrule
\end{tabular}}
\end{table}


\subsection{Case Study} \label{case_study}

Although both workflows demonstrate that three-second screenshots are sufficient for auto-journaling, differences arise between the feeling speculation and contextual analysis in the text-based versus video-based streams. In this section, we highlight key examples that showcase AutoJournaling's capabilities and the distinct internal logic between the two approaches.

\subsubsection{Case I: Event analysis}

\begin{figure}[ht]
  \centering
  \includegraphics[width=\linewidth]{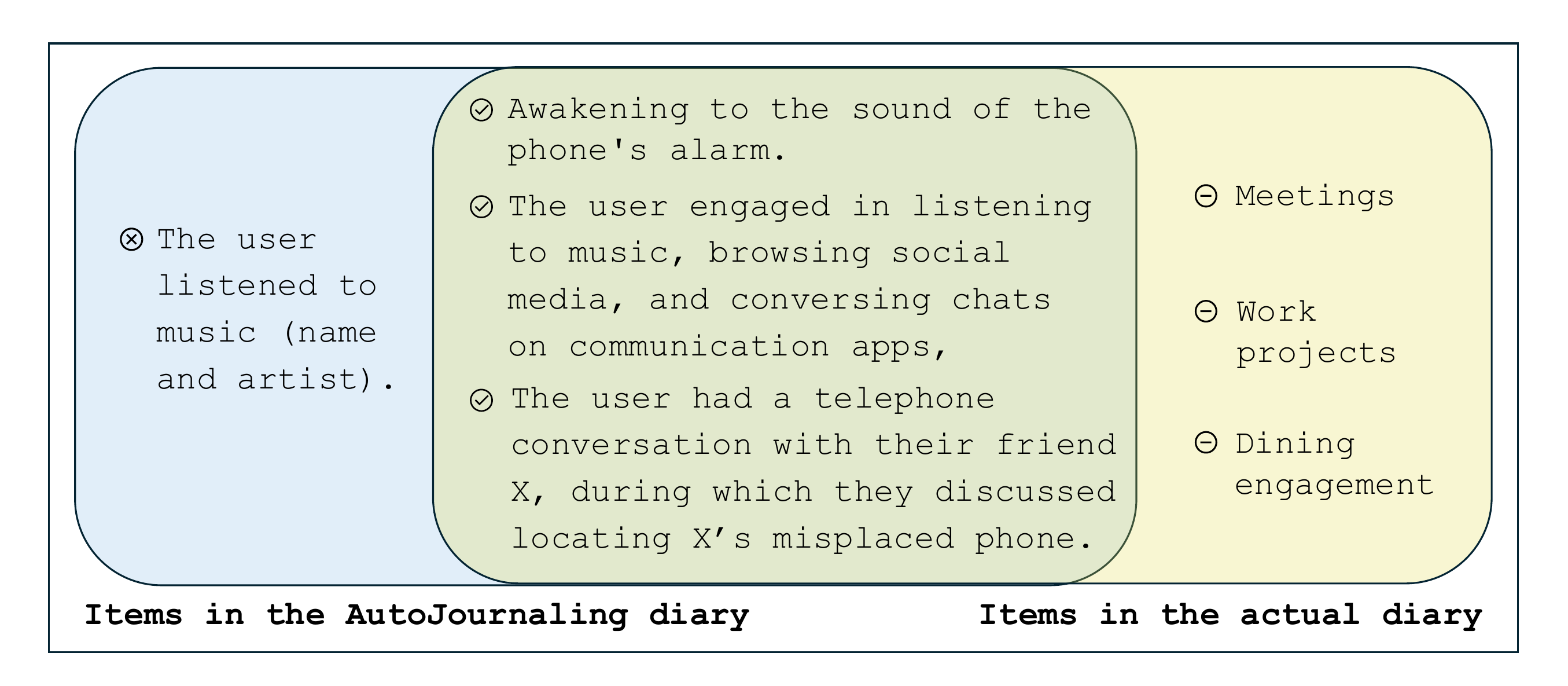}
  \captionsetup{font=footnotesize, labelformat=simple, labelsep=colon, labelfont=bf}
  \vspace{-10pt}
  \caption{An example comparison between the actual diary and the diary produced by AutoJournaling.}
  \label{diary_case}
\end{figure}

Comparing the events predicted by the AutoJournaling system and the ground truth recorded events, the AutoJournaling effectively captures and summarizes key events. In the case shown in Figure \ref{diary_case}, the significant events can be identified and summarized from chat histories in communication applications, telephony calls, smartphone alarms and music applications. 

As shown in Figure~\ref{diary_case}, the items not captured by the AutoJournaling system are primarily those related to participants' experiences that do not involve smartphone interactions and were, therefore, not expected to be detected by the system at this stage. For instance, records of in-person meetings, working contents or interactions during a dining event may not be captured unless explicitly mentioned in the smartphone data. While these elements are not currently included in the evaluation framework, future enhancements to the AutoJournaling system that incorporate other sensors could enable it to predict or infer such events beyond the screenshot. For example, geolocation sensing could help in determining travel events. Previous studies \cite{zhang2024leveraging, songyan2024leveraging} have explored this topic using smartphone-collected personal sensing data, demonstrating the strong performance of LLMs in handling such tasks. Furthermore, incorporating historical data into MLLMs and LLMs might enable the system to extract further personal insights from screenshots, improving the system's practicability for the predictions and monitoring of mental health issues.

While the system accurately generated most of the relevant items, it occasionally provided overly detailed information. For example, it sometimes included events such as listing specific music tracks, including their titles and artists, which were not mentioned in the participant's journal (Figure \ref{diary_case}). The level of detail desired in a journal is highly personal and can vary significantly between individuals. For this particular participant, such detailed entries were not part of their journal. Therefore, the AutoJournaling system could be further developed into a personalized model through contextual information, persona, and fine-tuning to better align with individual preferences.

\subsubsection{Case II: Feeling analysis}
The AutoJournaling system’s ability to infer feelings can be effective in general. Even when the exact words used to describe feelings differ, the system accurately captures the overall sentiments. For instance, as illustrated in Figure \ref{fig:event_feeling_example_video}, although "happy, connected" is not identical to "belonging, tired, and warm," the feelings still achieve a similarity score of 0.96. The detection of the event is particularly noteworthy given that no explicit references to kinship (e.g., words like "family", "mother", or "father") were present in the screenshots, highlighting the system’s nuanced understanding of context. Meanwhile, the tiredness mentioned in the participant's actual diary is not detected by the AutoJournaling system. This fatigue might be inferred from the participant's prior activities or sleep status. This indicates that though the system may perform well for event capturing, the analysis of feelings can be further improved by feeding more personalized and accurate contextual information.

Additionally, while the high evaluations scores (Table \ref{image_performance} and \ref{video_performance}) demonstrate the system’s capability to generalize emotional content, it also highlights that the system's reasoning process is more generalized than personalized. Since the screenshots did not include the content of the call, the emotional analysis was likely derived from either facial sentiment analysis (given that the call was a video call) or the model's prior knowledge. The reasoning suggests that the model assumes family calls generally convey positive sentiments without considering the specific information communicated during the call. Although this assumption matched the participant's experience in this case, it underscores the need for further exploration to understand and enhance the system's ability to provide more personalized emotional analysis.

\begin{figure}[htbp]
\centering
\fbox{
    \parbox{0.9\columnwidth}{
    \fontsize{8}{8}\selectfont
     \texttt{\textbf{Event}: Family call} \\
     \texttt{\textbf{Feelings}: Belonging, tired, warm.}
    }
}
\vspace{-6pt}
\captionsetup{font=footnotesize, labelformat=empty}
\caption*{(a) Actual Diary}
\vspace{4pt}
\fbox{
    \parbox{0.9\columnwidth}{
    \fontsize{8}{8}\selectfont    
     \texttt{\textbf{Event}: The user had a video call with their family.} \\
     \texttt{\textbf{Feelings}: Happy, connected} \\
     \texttt{\textbf{Reasoning}: The user is having a video call with their family, suggesting they are maintaining close relationships and enjoying a virtual connection.}
    }
}
\vspace{-6pt}
\captionsetup{font=footnotesize, labelformat=empty}
\caption*{(b) Diary Generated by the AutoJournaling Video-Based Stream}
\captionsetup{font=footnotesize, labelformat=simple, labelsep=colon, labelfont=bf}
\vspace{-4pt}
\caption{An example showing the event and feeling in the AutoJournaling-video-generated diary and the actual diary with the score of 0.93 for the event and 0.98 for the feelings.}
\label{fig:event_feeling_example_video}
\end{figure}

\vspace{-4pt}
\subsubsection{Case III: Contextual information and logic analysis}

In a case of one of the participants losing and finding their phone, the AutoJournaling system for the text-based analysis was able to meticulously document the entire event and infer the user’s emotions based on changes in circumstances as depicted in Figure \ref{fig:event_feeling_example_image}. The event is illustrated from the perspective of the lost phone, capturing key moments such as the activation of the `Find My Device’ function, a call from an unknown number, and the subsequent answer by a recipient. Furthermore, the emotional state `concerned' closely aligns with `anxious' and `helpless,' while when finding the phone `relieved' closely aligns with `extremely happy' and `relaxed,' effectively conveying the dynamic shifts in the user’s mood throughout the event.

However, compared to the text-based approach in AutoJournaling, the video-based approach can only capture segments of the events, often missing critical details and failing to provide a complete narrative. For instance, in the lost phone incident, the video streamline may not effectively categorize the important events into the lost phone event, such as messages inquiring about the lost phone, unknown calls, and the activation of the find-device service. Therefore, the text-based approach in the AutoJournaling system performs better than the video-based approach when the event unfolds over a longer duration. This advantage likely arises from the text-based approach’s ability to highlight detailed textual information, whereas the video-based approach tends to focus on the main events displayed on the smartphone, often failing to connect events occurring over a longer period into a cohesive narrative. To improve video analysis, it is crucial to enhance the AutoJournaling system's ability to recognize contextual information within a longer time window, ensuring that significant events are properly highlighted and contributing to a more comprehensive event diary.

\begin{figure}[htbp]
\centering
\fbox{
    \parbox{0.9\columnwidth}{
    \fontsize{8}{8}\selectfont
     \texttt{\textbf{Event}: Realized the phone was lost, used `Find my device' to locate the lost phone, asked a stranger to make a phone call, and then successfully got through the phone} \\
     \texttt{\textbf{Feelings}: Anxious, helpless; extremely happy, relaxed}
    }
}
\vspace{-6pt}
\captionsetup{font=footnotesize, labelformat=empty}
\caption*{(a) Actual Diary}
\vspace{4pt}
\fbox{
    \parbox{0.9\columnwidth}{
    \fontsize{8}{8}\selectfont    
     \texttt{\textbf{Event}: The user experienced a potential phone scare, using `Find My Device' to locate
    their phone, and then received and answered a call from an unknown number.}  \\
     \texttt{\textbf{Feelings}:  Concerned; relieved} \\
     \texttt{\textbf{Reasoning}:  Using `Find My Device' suggests concern about a lost phone. Locating
    the phone and answering a call implies a return to normalcy and relief.}
    }
}
\vspace{-6pt}
\captionsetup{font=footnotesize, labelformat=empty}
\caption*{(b) Diary Generated by the AutoJournaling Text-Based Stream}
\vspace{4pt}
\fbox{
    \parbox{0.9\columnwidth}{
    \fontsize{8}{8}\selectfont    
     \texttt{\textbf{Event}: The user is trying to locate their phone using the `Find My Device' feature on Google Play services.}  \\
     \texttt{\textbf{Feelings}:  Anxious, concerned}\\
     \texttt{\textbf{Reasoning}:  The user is actively trying to locate their phone, which suggests they might have misplaced it and are feeling anxious.}
    }
}
\vspace{-6pt}
\caption*{(c) Diary Generated by the AutoJournaling Video-Based Stream}
\captionsetup{font=footnotesize, labelformat=simple, labelsep=colon, labelfont=bf}
\vspace{-4pt}
\caption{An example showing the event and feeling in the AutoJournaling text/video-generated diary and the actual diary.}
\label{fig:event_feeling_example_image}
\end{figure}

\vspace{-8pt}
\subsection{Privacy and Limitations} \label{privacy}

While AutoJournaling offers automated self-reflection journaling and emotional state analysis for mental health and more generally psychological purposes, its method of capturing screenshots every three seconds and sending these images to cloud-based MLLMs like Gemini raises significant privacy concerns.
Continuous and frequent capturing of data compels users to expose their private smartphone activities, creating a sense of surveillance \citep{reeves2020time}. Moreover, uploading users' data to remote servers potentially leads to data leakage \citep{zhang2024enabling}. To address these privacy concerns, on-device LLMs \citep{zhang2024enabling} and private cloud computing \citep{AppleIntelligence} can be utilized to deliver personalized intelligent services. These technologies can be further explored to develop an on-device personalized AutoJournaling system, ensuring that user data remains secure while providing tailored journaling experiences.

The screenshots-to-journal methodology has limitations, including increased battery usage and storage demands. While the study confirms basic functionality, further research on larger datasets and cohorts is needed to enhance the system's ability to predict emotional states.

\section{Future Work and Conclusions}
In this paper, we have presented the AutoJournaling system, which captures screenshot information and leverages multi-modal large language models to automatically generate journals or narratives of smartphone users. Both the text and video stream methodologies we propose perform well in summarizing the journal events and inferring general psychological states. By analyzing screenshot data collected every 3 seconds from three university students over five days, we demonstrate the feasibility of the AutoJournaling system.

For future work, the AutoJournaling framework can be further enhanced in several ways, including the development of personalized models, the integration of on-device MLLMs, and combining it with screenshot collection systems \citep{reeves2021screenomics}. Since existing evaluation metrics are not designed for the tasks of comparing journal events and feelings, the evaluation process can be enhanced by incorporating human evaluations. Additionally, embedding temporal information in the journal generation and evaluation process could enable the model to predict the specific times at which events occurred, further refining utility of the generated journals. Also, this framework holds promise for predicting and intervening in problematic smartphone use, such as doomscrolling and social media addiction, by identifying issues through screenshot analysis without the need for annotated data.


\bibliographystyle{ACM-Reference-Format}
\bibliography{Reference.bib}

\end{document}